\documentclass[runningheads,a4paper]{llncs}

\usepackage{amssymb} 
\usepackage{graphicx}
\usepackage{url}
\urldef{\mailsa}\path|{m.khazeev, v.rivera, m.mazzara, l.johard}@innopolis.ru| 
\newcommand{\keywords}[1]{\par\addvspace\baselineskip
\noindent\keywordname\enspace\ignorespaces#1}

\begin{document}

% Eiffel commands
\newcommand{\eifkw}[1]{{\bf #1}}
\newcommand{\eif}[1]{\textsf{#1}}
\newcommand{\eifstring}[1]{''#1''}
\newcommand{\eifcomment}[1]{{-}{-}\textsf{#1}}
\newcommand{\eiftab}{~~~}
\newcommand{\eiftag}[1]{{\texttt{#1}}}

\mainmatter % start of an individual contribution

\title{Initial steps towards assessing \\ the usability of a verification tool}

%\titlerunning{Lecture Notes in Computer Science: Authors' Instructions}

\author{Mansur Khazeev \and Victor Rivera \and Manuel Mazzara \and Leonard Johard}
\authorrunning{}

\institute{Innopolis University, Institute of Technologies and Software Development,\\
1, Universitetskaya Str., Innopolis, Russia, 420500\\
\mailsa\\
\url{https://www.university.innopolis.ru}}


\toctitle{Lecture Notes in Computer Science}
\tocauthor{Authors' Instructions}
\maketitle

\begin{abstract}
In this paper we report the experience of using AutoProof for static verification of a small object oriented program. We identify the problems that emerge by this activity and classify them according to their nature. In particular, we distinguish between tool-related and methodology-related issues, and propose necessary changes to simplify both the tool and the method.
\keywords{Static Verification, AutoProof, Verification issues}
\end{abstract}

\section{Introduction}
\label{sec:intro}
%\todo{Bertran: Show 80-90\% of results}

Formal proof of correctness of software is still not commonly accepted in practice, even though both hardware and software technologies for verification have significantly improved since it was first mentioned in the context of "verifying compiler"\footnote{A cipher for an integrated set of tools checking correctness in a broad sense \cite{vercompiler,tokexperiment}}. In ideal world verifying software would need only ``pushing a button'', though this kind of provers exist, but they are limited to verification of simple or implicit properties such as absence of invalid pointer dereference \cite{modelchecking:clark:2000}. In order to verify a software, a formal specification should be provided against which it will be verified. Given a specification, like contracts in Design-by-Contract (DbC) methodology, it is possible to verify specific implementations with respect to this specification. The term Design-by-Contract was originally introduced in connection with the design of the Eiffel programming language, but is nowadays also adopted in many other languages. For example, in C\# the methodology is supported through an additional library \cite{cContracts}. Java has JML add-on \cite{Leavens03designby}, while Kotlin has preconditions (\textbf{require} and \textbf{requireNotNull} clauses) implemented at the language level. Contracts are fully supported in Eiffel. 

Eiffel has a prover for functional correctness called AutoProof \cite{autoproof:julian:15}. This prover comes with a powerful methodology for framing and class invariants and it fully supports advanced object-oriented features \cite{semcol:Sem2014}. We here present a series of case studies in order to test the usability of the tool and its applicability in general practice . The tool was used for verification of three exercises of different size and complexity: a simple class, a set of related classes and small size industrial project. This paper describes the results of the first exercise - verification of the class SET, that implements classic sets from set theory: properties and classical operations.

The challenge of this exercise is mainly related to difficulties that a new user can encounter while using the tool for the first time. There is no explicit documentation available: only the website and several papers from the authors of the tool as the main source of information. However the notation has been evolving and in some of these papers it is no longer relevant. Verification with AutoProof often requires additional annotation that helps the tool to derive the more “complex” properties from the trivial ones. However, for someone who does not know how the tool works and what is going on under the hood, the feedback from the tool can be useless or even confusing. Naturally, this might be excusable if the tool is meant to be used by a limited group of scientists, but complete documentation needs to be developed, thereby minimizing the need of additional assertions, in order to make a verification tool applicable in industrial practice. This is essential, because the tool still requires a knowledge of the underlying mechanisms and a number of additional annotations.

\section{Eiffel and Autoproof}
Eiffel is an object oriented programming language that natively supports the Design-by-Contract methodology \cite{autoproof:julian:15}. All features in Eiffel should be specified through equipping them with contracts, namely pre- and post-conditions; as well as properties of classes through invariants. AutoProof is a static verifier for programs written in Eiffel. It  follows the auto-active paradigm\cite{autoproof:julian:15} where verification is done completely automated, similar to model checking \cite{modelchecking:clark:2000}, but where users are expected to feed the classes providing additional information in the form of annotations to help the proof. The tool is capable of identifying software issues without executing the code, thereby opening a new frontier for ``static debugging'', software verification and reliability in addition to general improvements to software quality.

AutoProof verifies the functional correctness of a code written in Eiffel language equipped with contracts. The tool checks that routines satisfy pre- and post-conditions, maintenance of class invariants, loops and recursive calls termination, integer overflow and non \eifkw{Void} (i.e. $null$ in many other programming languages) references calls. For that purpose, AutoProof uses a verification language called Boogie \cite{Rustan:Boogie:08}: AutoProof translates  Eiffel code into Boogie programs as an intermediary step. The Boogie tool generates verification conditions (logic formulas whose validity entails the correctness of the input programs) that are then passed to an SMT solver Z3. Finally, the verification output is returned to Eiffel.

AutoProof supports most of the Eiffel language constructs: in-lined assertions such as \eifkw{check} ($assert$ in many other programming languages), types, multi-inheritance, polymorphism. By default AutoProof only verifies user-written classes when a program is verified, while referenced libraries should be verified separately or should be based on pre-verified libraries, e.g. \texttt{EiffelBase2} \cite{PTF-FM15}. This pre-verified library offers many different data structures with all features fully verified.

\section{Case study experience}
The first stage in series of case studies was verification of simple example -- the implementation of an ordinary class for a generic implementation of sets, \eif{MY\_SET}, using lists ( \eif{V\_LINKED\_LIST} from the \texttt{EiffelBase2} library) and equipping it with contracts. Corresponding annotations were added to help AutoProof to prove the class.

Set properties were expressed as invariants, namely:
\begin{itemize}
\item No duplicate elements
\item Order of elements in the set is not important
\item Cardinality is always greater or equal to 0
\end{itemize} 

The class implements some basic set operations:
\begin{itemize}
\item \eif{is\_empty} - a query that states whether the set contain no elements
\item \eif{cardinality} - number of elements in the set
\item \eif{has} - a query that states whether the set contains a given element
\item \eif{is\_strict\_subset}, \eif{is\_ subset} - queries that states whether the set is a subset (or a strict subset) of a given set
\item \eif{union}, \eif{intersection}, \eif{difference} - functions returning new set with the union, intersection of difference with a given set, respectively
\end{itemize}

%Vic: not sure about the following paragraph. Mansur, can you elaborate more?
During the verification process, it turned out, that working with V\_ classes was too complicated for non-expert users. Therefore the decision was done to simplify the example replacing V\_LINKED\_LIST with SIMPLE\_LIST.

%{\color{red}TODO: Explain how big the code is

%TODO: The "story" (process) of verification}

\section{Problems taxonomy}

%\subsection{Misunderstandings}
%\subsubsection{}

Despite the simplicity of the class, various problems arose due to lack of user experience with the AutoProof tool, ranging from issues with the tool installation all the way to issues with checking the verified class with tests. In our analysis, these problems have been divided into two main categories: problems with the tool and problems with the approaches or methodologies used in the tool.

\begin{enumerate}
\item Problems with the tool
\begin{multicols}{2}
  \begin{enumerate}
  \item Lack of documentation
  \item Poor tool feedback
  \item Redundancy in notations
  \item Misleading notations
  \item Order of assertions
  \item Limitations of the tool
  \item User Interface (UI) bug
  \item Difficulties with installation/compilation from the sources
  \end{enumerate}
  \end{multicols}
\item Problems with methodologies
\begin{multicols}{2}
  \begin{enumerate}
  	\item Semantic collaboration
    \item Framing
  \end{enumerate}
    \end{multicols}
\end{enumerate}
The first category includes rather minor problems and bugs, mostly related to the particular implementation in the tool and means that those require local fixes. However, the second category require improving the methodology or replacing them with the alternative ones.

\subsection{Problems with the tool}

\begin{figure}[!ht]
\centering\includegraphics[width=.8
\linewidth]{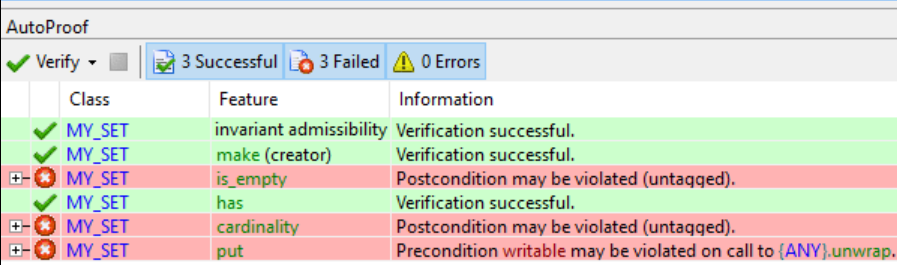}
\caption{UI of AutoProof}
\label{fig:ui}
\end{figure}

The challenge of this exercise was mainly related to the fact that it had to be done by a person who has no previous experience with AutoProof, nor any other similar tools. The difficulty is not in some sophisticated user interface (UI), quite the opposite, it is rather simple (see figure \ref{fig:ui}) - a ``Verify'' button and a table, where the results are being displayed. The main obstacle is in the fact that, the tool expects an input in terms of assertions, and it is not always clear what the real effect of each input is.

\subsubsection{Lack of documentation}
As we previously described, the tool requires additional annotations that assist the verification and help to derive one property from another. Although AutoProof exploits the syntax of the Eiffel language, additional annotations have been introduced by the developers of the tool. Most of them are briefly described in the online manual, which is available on the EVE website\footnote{EVE (Eiffel Verification Environment). EVE is a development environment integrating formal verification with the standard tools}. In addition, there is an online tutorial which is useful for quick acquaintance with the tool. However, this is clearly an insufficient reference material for working with the tool.

Overall, there is not much of documentation available online: the website, and several papers from the authors of the AutoProof tool.Moreover, the notation has been evolving and in some of these papers it is no longer relevant.Naturally, this might be excusable if the tool is meant to be used by a limited group of scientists. On the other hand, if the idea is to apply verification in industrial practices,  then documentation is essential. The tool still requires documentation that explains the annotations needed and the knowledge of the different mechanisms.

\subsubsection{Poor tool feedback}
The process verification (or static debugging) starts with pushing a ``Verify'' button. The tool then returns some feedback in the form of success, error and failure messages. A failure message is a message that shows the property that cannot be proved. An error message consists of information on some issue with the input. In both case, whether it is error or failure messages, users need to fix them by adding missing assertions. This process repeats until the class is fully verified. AutoProof implements a collection of heuristics known as ``two-step verification'' that helps discern between failed verification due to real errors and failures due to insufficiently detailed annotations \cite{autoproof:julian:15}. These failure messages are usually informative: they describe the property and sometimes the reason of the failure. On the other hand, error messages usually do not tell more than that the tool cannot proceed with the input it has received. 

If there is an error during the translation to Boogie the verification process stops and AutoProof returns an error message about ``internal failure'' in some cases with no additional information. Usually, this errors are caused by newly entered assertions, which makes the process of correcting them easier. However, if this is not the case, then it is difficult to understand what exactly causes the error. This may become an issue when verifying the whole class, with all its implemented features and stated contracts, because it is not possible to determine the source of the error. The solution might be to comment out the features and iteratively verify them one by one by decommenting them.

\subsubsection{Redundancy in annotations}
AutoProof supports most of the Eiffel language as used in practice \cite{autoproof:julian:15}. It also introduces some new notations that support the methodologies used for verification. These notations are useful for manipulating the defaults of semantic collaboration in features and classes (this will be discussed in \ref{sec:semcollaboration}).

However, some of these additional notations introduce redundancy. For example all creation procedures in Eiffel must be listed under the key word \eifkw{create}. Autoproof does not make use of this and instead expects the user to explicitly declare the creation procedure as depicted in Figure \ref{fig:creation}. Even though the procedure \eifkw{make} is defined as a creation procedure in Eiffel, the verifier expects an additional \eifkw{note} clause with \eiftag{status:} \eifkw{creator} in order to treat it as creation procedure.

\begin{figure}[!ht]
{\small
\[	
\begin{array}{l}
\eifkw{create}~\eif{make}\\
\eifkw{feature}~\\
\hspace{0.2cm}\eif{make}\\
\hspace{.4cm}\eifkw{note}\\
\hspace{.8cm}\eiftag{status: } \eif{creator}\\
\hspace{.4cm}\eifkw{do}~\eif{\dots}~\eifkw{end}\\
\end{array}
\]
} 
\caption{Accepted creation procedure by Autoproof.}
\label{fig:creation}
\end{figure}

Another example is the possible inconsistencies on the given annotations. In Autoproof, one can declare a procedure as \eifkw{pure}, specifying that it will not change the state of the object, or \eifkw{impure} specifying that the procedure might change the state. This can also be achieved by listing the locations that the procedure might change. This is done by using the annotation \eifkw{modify}. If the clause is empty it means that the function is \eifkw{pure}, \eifkw{impure} otherwise. For Autoproof to be able to prove the procedure \eifkw{union} in Figure \ref{fig:impure}, it has to be defined with the \eifkw{impure} annotation. This means that it does modify the state of the object. The empty clause \eifkw{modify} then needs to specify that the function is pure. This is done in order to be able to use \eifkw{wrap} and \eifkw{unwrap} in the function (as explained later on). 

\subsubsection{Misleading notations}
AutoProof support inline assertions and assumptions, which can be expressed using the \eifkw{check} clause supported by Eiffel. \eifkw{Check}s are intermediate assertions that are used during the debugging process in order to check whether the user has the same understanding of the state at a program point as the verifier \cite{tutorial}. However, removing an intermediate \eifkw{check} clause from successfully verified feature might make the verification process fail. This, more than being there for the user, is due to \eifkw{check} assertions and guides the verifier towards a successful verification. Probably this is a design solution: not to introduce another clause but to use an existing one from the language. However, this might confuse users.

\subsubsection{Order of assertions}
A \eifkw{check} clause is useful because the verifier does not just check the property enclosed, but also uses it for further derivations in case the property was proved correct. The same applies to class invariants, and that makes the order of properties substantial for the tool. This means that properties are joint not by the \eifkw{and} operator, but by \eifkw{and then}, which may lead to verification failures even if all needed properties are stated (although in improper order). For example, there are two assertions depicted in figure \ref{fig:order}: the first for setting up the relation between elements (the model) and data (the implementation); the second, for defining owned object by \textbf{Current}\footnote{denoting the current object in Eiffel}. However, in this order the verification will fail, while it will succeed if \textbf{owns\_def} is stated first.

\begin{figure}[!ht]
{\small
\[	
\begin{array}{l}
\eifkw{invariant}~\\
\hspace{.4cm}\eifkw{model\_def: elements = data.sequence.range}\\
\hspace{.4cm}\eifkw{owns\_def: owns = [data]}\\
\hspace{.8cm}\eif{\dots}\\
\end{array}
\]
} 
\caption{The order of invariant assertions.}
\label{fig:order}
\end{figure}

\subsubsection{Limitations of the tool}
Null pointer dereferencing is a well-known issue in object-oriented programming. In Eiffel, this can be avoided by letting the compiler check for call consistency \cite{kogtenkov:2017}: the object source making the call cannot be a Void object. Currently, Autoproof does not make use of this property of Eiffel. For instance the verified library \texttt{EiffelBase2} can only be used when the void-safety property of Eiffel is disabled. There is a coming version of the tool to support these two Eiffel environments, but the version is not available yet.

\subsubsection{User Interface (UI) bug}
The tool lacks support which can be observed in some rare bugs. For example, it can skip some of the features of the class or verify only one of the features instead of the whole class. Even though, the tool never returned improper successful verification results, these kinds of bugs might be disrupting to the user.

\subsubsection{Difficulties with installation/compilation from sources}
There are two ways to get the tool working on a local computer: by installing the build (available online) or compiling the tool from the source code. For the latter option, the repository requires a  clean-up for compiling. Therefore, is better to use the former method.

In addition, there are several manipulation has to be done while creating a new project in AutoProof, such as disabling some options and reopening the project in order to clean it. 

\subsection{Problems with methodology: Semantic collaboration and Framing}
AutoProof supports advanced object-oriented features through a powerful methodology to specify and reason about class invariants of sequential programs \cite{semcol:Sem2014}. But this power comes at the price of simplicity - the tool requires users to understand all the underlying methodologies. This limits the tool to expert users by exceedingly complicating the verification of even such simple classes.

\subsubsection{Semantic collaboration}
\label{sec:semcollaboration}
AutoProof supports semantic collaboration, i.e. the full-fledged framing methodology
that was designed to reason about class invariants of structures made of collaborating objects \cite{semcol:Sem2014}. This methodology introduces its own annotations which do not exist in the Eiffel language. Annotations are used to equip features and entire classes with additional information which are used by the verifier. These include \eif{ghost} attributes -- class members used only in specifications -- which are useful when maintenance of global consistency is required as in subject/observer or iterator pattern examples \cite{semcol:Sem2014}. These ghost attributes and default assertions that are added into pre- and post-conditions often result in over-complicating the verification process of rather simple classes. 

During initial steps of the verification process of the case study presented in this paper, time was spent on trying to understand the failure message: ``\textbf{default\_is\_closed} may be violated on calling some feature'' for some private attribute. Basically, the tool was expecting \eif{owns = [data]} in the invariants of the class which is not obvious without understanding the methodology. Moreover, for this specific example the property could have been derived from exportation status of the attribute. Eiffel language supports the notion of ``selective export'', which exports the features that follow to the specific classes and their descendants \cite{Meyer:Touch:2009}. The verifier ignores this useful information and requires the properties to be stated explicitly. Considering selective export might decrease the need for using semantic collaboration \cite{expstat}. 

\subsubsection{Framing}
The framing model is used in AutoProof in order to help reason about objects that are allowed/not allowed to be updated. There are different ways to specify this, for instance by adding modifies clauses in pre-conditions. One can specify one or more model fields, attributes of the class or list of objects which may be updated. This is rather intuitive and straightforward, though it seems to be more relevant to post-condition clauses. Another alternative is to make use of default clauses included into each routine, so the framing model should be used only if the behavior of the routines is different than from default. For example, in \eif{MY\_SET} class, all routines are pure (no side effects), hence all routines were equipped with an empty  \eif{modify} clause. Even in a function that is defined as pure using the \eif{modify} clause, that function needs to be specified as impure in order to use \eif{is\_wrapped} clause, even though it does not modify the state of any object (see figure \ref{fig:impure}). This might confuse the user.

%The other thing was related to \textbf{Reads ()} clause. It was not clear why one need to specify the fact that a routine is "allowed to read the specified objects, attributes or models".

%\subsubsection{Model queries}

\section{Related Work}
\label{sec:related}

Formal notations to specify and verify software systems have existed for a long time, in particular in some specific domain such as process modeling \cite{YanMCU07}. A survey of the major approaches can be found in \cite{Mazzara10}, while \cite{Mazzara09} discusses the most common methodological issues of such approaches. Another approach, as in \cite{rivera:eb2java:14,Rivera:2017}, is to use the formal notation of a modeling language to specify and verify software systems to then translate it to a programming language. 

\begin{figure}[!ht]
{\small
\[	
\begin{array}{l}
\eifkw{feature}~ \eifcomment{ Queries}\\
\hspace{0.3cm}\eif{union}(\eif{other}: \eifkw{ like ~Current }): \eifkw{ like ~Current }\\
\hspace{.6cm}\eifcomment{ New set of values contained in `Current' or `other'}\\
\hspace{.4cm}\eifkw{note}~
\eif{status}:\eif{impure}\\
\hspace{.4cm}\eifkw{require}\\
\hspace{.6cm}\eif{ modify\_nothing }: \eif{ modify([ ])}\\
\hspace{.8cm}\eif{\ldots}\\
\hspace{.4cm}\eifkw{end}
%\hspace{.4cm}\eifkw{do}\\
%\hspace{.8cm}\eif{...}\\
%\hspace{.4cm}\eifkw{end}\\
\end{array}
\]
} 
\caption{Pure function (empty \eif{modify} clause) specified as impure (note clause \eiftag{status:} \eif{impure})}
\label{fig:impure}
\end{figure}

In \cite{DSilvaKW08} the authors present an extensive survey of algorithms for automatic static analysis of software. The discussed techniques (static analysis with abstract domains, model checking, and bounded model checking) are different, but complementary, to the one discussed in this paper, and they are also able to detect programming errors or prove their absence. 

The importance of focusing on usability requirements for verification tools has been identified in \cite{Rozilawati10}. The authors have classified usability properties into three main categories: Interface, Utility and Resources management. Since the interface of AutoProof tool consists of a button and a table, the interface category was omitted. Only utility (in term of clearness of error/failure messages)  and Resources management (in term of properties such as installation, documentation) were considered.

The results of testing the usability of AutoProof, in particular, by non-expert users has been studied in \cite{FPT-FIDE15}, where programmers with little formal methods experience were exposed to the tool.

%the pros of both rigorous methodologies and supporting tools able to semi-automate the process. Before this to be available for the average developer it is however necessary to improve the users' experience. A comparison between different approaches (for example Event-b/Rodin and Design-by-contract/Autoproof) is beyond the scope of this paper and it is left as future work. 
%{\color{red}TODO: Write the related work section based on last year paper}

\section{Conclusion}
\label{sec:conclusion}
AutoProof is not trivial in its usage and needs detailed knowledge of what is going on behind the scenes. The tool requires a number of additional assertions in pre- and post-conditions, as well as in invariants for successful verification, while ignoring some information that has been already provided. To be used in practice the usability of the tools needs to be be significantly improved to the level where verification is “simple” enough to be used by ordinary programmers. By “simple” we mean, that it should:
\begin{itemize}
\item require less additional annotations by automatically deriving properties from information which is currently being neglected and by removing redundant clauses and reworking some of ghost class members;
\item provide clearer feedback in case some property can not be satisfied, offering hints and possible solutions;
\end{itemize}
In addition, it is important to:
\begin{itemize}
\item develop a documentation describing all used methodologies, including detailed information about notations with examples \item clean up and rebuild the tool from latest sourced that are available in the EVE repository and fix all the bugs that we identified;
\end{itemize}
As a further work, AutoProof will be tested through verification of a set of related classes and a small size industrial project, the Tokeneer project\footnote{\url{http://www.adacore.com/sparkpro/tokeneer/download}}.

\bibliographystyle{ieeetr}
\bibliography{references}

\begin{thebibliography}{10}

\bibitem{vercompiler}
J.~King., {\em A Program Verifier}.
\newblock PhD thesis, School of Computer Science, Carnegie Mellon University,
  1969.

\bibitem{tokexperiment}
J.~Woodcock, E.~G. Aydal, and R.~Chapman, {\em The Tokeneer Experiments},
  pp.~405--430.
\newblock 2010.

\bibitem{modelchecking:clark:2000}
E.~M. Clarke, Jr., O.~Grumberg, and D.~A. Peled, {\em Model Checking}.
\newblock Cambridge, MA, USA: MIT Press, 1999.

\bibitem{cContracts}
M.~Documentation, ``Code contracts.''
  https://msdn.microsoft.com/en-us/library/dd264808, accessed in May 2017.

\bibitem{Leavens03designby}
G.~T. Leavens and Y.~Cheon, ``Design by contract with jml,'' 2003.

\bibitem{autoproof:julian:15}
J.~Tschannen, C.~A. Furia, M.~Nordio, and N.~Polikarpova, ``Autoproof:
  Auto-active functional verification of object-oriented programs,'' in {\em
  21st International Conference, {TACAS} 2015, London, UK, April 11-18, 2015.
  Proceedings}, pp.~566--580, 2015.

\bibitem{semcol:Sem2014}
N.~Polikarpova, J.~Tschannen, C.~A. Furia, and B.~Meyer, {\em FM 2014: Formal
  Methods: 19th International Symposium, Singapore, May 12-16, 2014.
  Proceedings}, ch.~Flexible Invariants through Semantic Collaboration,
  pp.~514--530.
\newblock Springer International Publishing, 2014.

\bibitem{Rustan:Boogie:08}
K.~R.~M. Leino, ``This is boogie 2,'' tech. rep., June 2008.

\bibitem{PTF-FM15}
N.~Polikarpova, J.~Tschannen, and C.~A. Furia, ``A fully verified container
  library,'' in {\em FM 2015: Formal Methods}, Lecture Notes in Computer
  Science, Springer, 2015.

\bibitem{tutorial}
E.~Z. Chair~of Software~Engineering, ``Autoproof tutorial,''

\bibitem{kogtenkov:2017}
A.~Kogtenkov, {\em Void Safety}.
\newblock PhD thesis, ETH Zurich, 2017.

\bibitem{Meyer:Touch:2009}
B.~Meyer, {\em Touch of Class: Learning to Program Well with Objects and
  Contracts}.
\newblock Springer Publishing Company, Incorporated, 1~ed., 2009.

\bibitem{expstat}
D.~de~Carvalho, ``Modularly reasoning in object-oriented programming using
  export status.'' unpublished, 2017.

\bibitem{YanMCU07}
Z.~Yan, M.~Mazzara, E.~Cimpian, and A.~Urbanec, ``Business process modeling:
  Classifications and perspectives,'' in {\em Business Process and Services
  Computing: 1st International Working Conference on Business Process and
  Services Computing, {BPSC} 2007, September 25-26, 2007, Leipzig, Germany.},
  p.~222, 2007.

\bibitem{Mazzara10}
M.~Mazzara and A.~Bhattacharyya, ``On modelling and analysis of dynamic
  reconfiguration of dependable real-time systems,'' in {\em 2010 Third
  International Conference on Dependability}, pp.~173--181, July 2010.

\bibitem{Mazzara09}
M.~Mazzara, ``Deriving specifications of dependable systems: toward a method,''
  in {\em Proceedings of the 12th European Workshop on Dependable Computing,
  EWDC}, 2009.

\bibitem{rivera:eb2java:14}
V.~Rivera and N.~Cata{\~n}o, ``{Translating Event-B to JML-Specified Java
  programs},'' in {\em 29th ACM SAC}, (Gyeongju, South Korea), March 24-28
  2014.

\bibitem{Rivera:2017}
V.~Rivera, N.~Cata\~{n}o, T.~Wahls, and C.~Rueda, ``Code generation for
  event-b,'' {\em Int. J. Softw. Tools Technol. Transf.}, vol.~19, pp.~31--52,
  Feb. 2017.

\bibitem{DSilvaKW08}
V.~D'Silva, D.~Kroening, and G.~Weissenbacher, ``A survey of automated
  techniques for formal software verification,'' {\em IEEE Trans. on CAD of
  Integrated Circuits and Systems}, vol.~27, no.~7, pp.~1165--1178, 2008.

\bibitem{Rozilawati10}
R.~Razali and P.~Garratt, ``Usability requirements of formal verification
  tools: A survey,'' {\em Journal of Computer Science}, vol.~10, no.~6,
  pp.~1189--1198, 2010.

\bibitem{FPT-FIDE15}
C.~A. Furia, C.~M. Poskitt, and J.~Tschannen, ``The {AutoProof} verifier:
  Usability by non-experts and on standard code,'' in {\em Proceedings of the
  2nd Workshop on Formal Integrated Development Environment (F-IDE)}
  (C.~Dubois, P.~Masci, and D.~Mery, eds.), vol.~187, pp.~42--55, EPTCS, June
  2015.

\end{thebibliography}

\end{document}